\documentstyle[12pt]{article}
\begin{document}
\begin{center}
{\large \bf Collective Political Opinion Formation in Nonlinear
Social Interaction
\\

\vspace*{.5in}

\normalsize
Soo Yong Kim$^{1}$, Chung Hyun Park$^{1}$,
Kyungsik Kim$^{2,*}$  \\

\vspace*{.1in}

{\em
$^{1}$Department of Physics, Korea Advanced Institute\\
of Science and Technology, Daejeon 305-701, Korea\\
$^{2}$Department of Physics, Pukyong National University,\\
Pusan 608-737, Korea\\
} }
\end{center}

\hfill\\
\baselineskip 24pt
\begin{center}
{\bf Abstract}
\end{center}
\hfill\\
We have presented a numerical model of a collective opinion
formation procedure to explain political phenomena such as two-party
and multi-party systems in politics, political unrest, military coup
d'etats and netizen revolutions. Nonlinear interaction with binary
and independent decision making processes can yield various
collective behaviors or collective political opinions. Statistical
physics and nonlinear dynamics may provide useful tools to study
various socio-political dynamics.

\vskip 10mm
\hfill\\
PACS number(s): 89.65.Gh, 02.50.Ey, 05.40.Jc, 05.45.Tp \\
$Keywords$: Social Interaction, Periodic oscillation, Binary choice,
Political party, Bifurcation map

\vskip 1mm
\hfill\\
Corresponding author. Tel.: +82-51-620-6354; Fax: +82-51-611-6357\\
$^{*}Email Address$: kskim@pknu.ac.kr(K. Kim) \\

\newpage

\noindent
{\bf 1. INTRODUCTION}
\hfill\\

In social sciences, such as sociology, demography, regional sciences
and economics, nonlinear social interaction between human
individuals has been analyzed in detail $[1-8]$. Some examples
include collective behaviors, collective political opinion
formation, emergence of social dilemmas, regional migration of
interactive populations, settlement formation on the various spatial
scales, and market instability in non-equilibrium economics.
Collective decision making procedures and policy determination
produce unstable and unpredictable outcomes, referred to as
nonlinear and chaotic behavior, in a socio-dynamic system. Evidence
of this is readily found in the field of financial markets.  As the
simplest model to explain collective opinion formation, it is quite
natural that a large class of these individual decisions may be
binary, which is similar to McCulloch-Pitts' neuron $[9]$, from a
sociological perspective on individual behavior. These binary
strategies or choices are basic ingredients to determine an
individual's character, the stability or health of a society, and
even a financial market's efficiency.

Opinion formation of individuals within their own political systems
is a typical example that can be examined and investigated. A
political party is a political organization that subscribes to a
certain ideology and seeks to attain political power within a
government. The party's policies often represent an aggregation of
interests within the party. These interests will inevitably vary
considerably, even amongst party members. The simple case of two
political opinions corresponds to two-party systems, for example,
the Democratic and the Republican parties in the USA, where these
two political parties are dominant. In contrast, the Italian
political system is a multi-party system. In a multi-party system,
there may exist two strong parties, with a third party that is
electorally successful, as found in the UK and Canada. The party may
frequently come in second place in elections and pose a threat to
the other two parties, but has still never held government formally.
Finland may have an active three-party system, in which all three
parties hold top office. It is very rare for a country to have more
than three parties who are all equally successful, and all have an
equal chance of independently forming government. In cases where
there are numerous parties, no one party often has a chance of
gaining power, and parties must work with each other to form
coalition goverments. Sometimes existence of multi-parties lead to a
political spectrum where the left is associated with radical or
progressive policies and the right with conservative policies.

Statistical physics and synergetics provide us with a powerful tool
to examine and study social phenomena with nonlinear interaction,
both qualitatively and quantitatively. The physical scheme wherein
the behavior of an atom is influenced by the presence of other atoms
is similar to the assumption within social science that one person's
decisions depend on the decisions of the others. The process of
going from one atom to many atoms in bulk has much in common with
the process of one individual proceeding to a social group or
organization. Physics has been successful in describing macroscopic
behavior using microscopic variables, and thus the somewhat
ambitious application of statistical physics to complex social
phenomena based on well-established physical formalism is not
illogical.\\

\noindent {\bf 2. Models }
\hfill\\

Let us consider a model of binary decisions made by a separate group
of individuals who maintain a common interest. The collective
behavior of people deciding to vote for or against a political party
can be modeled. Each individual can make a choice w. w can be coded
as 0(against) or 1(for), which can be viewed as two elements in
Iching $[10]$, i.e., yin and yang. Yin and yang are a philosophical
concept, a means to generalize two opposite principles that may be
observed in nature. The terms yin and yang are applied to express
dual and opposite qualities, for example, day and night, brightness
and dimness, movement and stillness, heat and cold, upward and
downward directions, etc.

The theory of yin-yang holds that everything in nature has two
opposite aspects that are mainly reflected in their ability to
struggle with, and thus control each other. Warmth and heat can
repel coldness, while coldness may lower a high temperature. The yin
or yang within any phenomena will restrict each other through
competition and opposition, which can be a binary decision made by
our God in nature. In China, Japan, and Korea, it is widely believed
that the relative physiological balance in the human body will be
destroyed and disease will arise if for any reason this mutual
opposition results in an excess or deficiency of yin or yang.

In order to describe this dichotomy between for and against, we
employed a fashion model with social interaction $[11]$ in this
study and modified the utility function as
\begin{equation}
V=\beta x^{\alpha} (1-x)^{\gamma}-b, \label{eq:a1}
\end{equation}
where $\alpha$ or $\gamma$ are control parameters to represent the
support index of "for' or 'against' opinions. $\beta$ is a coupling
constant that is related to a social temperature $T$, $b$ is a
threshold that determines the health state of a social system, and
$x$ is the rate of binary choice, for example pros or cons
($0<x<1$). The dynamical system of binary choice can then be
expressed as
\begin{equation}
x_{i+1} =\frac{1}{1+\exp[-\beta{x_i}^{\alpha} (1-x_i )^{\gamma}+b]}
\label{eq:b2}
\end{equation}
In this equation, it is assumed that the process of creation or
transformation of political opinion can be modeled as a flow of
social interactions between people with different opinions. Due to
the distribution of political opinion at any moment and of the
permeability of the people to social interaction, the individuals
favor and support a specific opinion. This model may be applied to
collective political opinion formation and it is examined through
numerical experiments.\\

\noindent {\bf 3. Numerical results }
\hfill\\

The model behaviors, which are used to deduce the decision-making
procedure of collective opinion in a society, are classified into
three categories, stable convergence, stable periodic oscillation,
and unstable periodic oscillations leading to chaos. They can be
interpreted in terms of a unification of single dominant group
opinion, two or three definite but rival opinions in conjunction
with period, and a chaotic or uncontrollable state mixed with
numerous rival opinions.

Fig. $1$ shows a bifurcation map for stable convergence or
bi-stability, such as two definite 'for' or 'against' opinions with
$\alpha=0.2$, $\gamma=0.8$, $b=3$ and a varying constant $\beta$.
For $\beta < 6.42$, i.e., a high temperature region, the curve
converges to a fixed point. This means that a single and unified
opinion of individuals may emerge within their own political
systems. When $6.42 < \beta$, i.e., a low social temperature region,
there are two split curves, and the iterates of all $x$ are
attracted to the $2$-period cycle. This corresponds to a stability
of two independent and concrete political opinions under social
temperature low enough to excite two modes. There are two parties,
such as the majority and the minority party in a democratic
political system, for example long-lasting Democratic or Republican
parties in the American political system. No other collective
opinions except two independent ones or emergent behavior of two
modes can be possible in this society, which may be regarded as
politically stable and healthy due to various public opinions
influenced by the mass media, the Internet, and government or
non-government organizations.

Fig. $2$ also shows a bifurcation map for stable convergence,
bi-stability, tri-stability, and chaotic behavior with $\alpha=0.5$,
$\gamma=0.5$, $b=3$ and varying social temperature $\beta$. For
$\beta < 9.35$, i.e., a very high social temperature region, the
curve converges to a fixed point again. When $\beta=9.35$, two
curves start to appear again, and the iterates of all $x$ are
attracted to the $2$-period cycle, such as majority and minority
opinions. The two curves are subsequently divided into $4$ curves
and the iterates of all $x$ are attracted to the $4$-period cycle. A
single majority opinion is split into two majority opinions, which
can eventually be unified back into a single majority opinion in
some cases. The single minority opinion is also split into two
opinions. A window with a $3$-period cycle is also detected at
approximately ¥ã=18.5 on the upper tongue of the bifurcation diagram
(inset in Fig. $2$). There exists a single intermediate opinion, a
so-called third party's opinion, in addition to both the majority
and minority opinions.
The collective opinion curve can also be detected as periodic or
chaotic as an intermittency route in various social temperatures,
leading to political unrest. The chaotic behavior of collective
opinions represents political chaos and unrest due to revolutionary
breakdown, war, financial crisis, or terror. The former apparent
eternity and sudden collapse of eastern communist leaderships can be
explained through this political unrest.

One stable opinion, two rival opinions, four different opinions,
three opinions, and chaotic states of many different opinions occur
with varying social temperature. Occurrence of chaotic states is
eventually accompanied by social development, harmony and action to
equally double political opinions. There is no cosmos without chaos.

The periodic or chaotic opinions depend on underlying political
dynamics controlled by various control parameters, $\alpha$,
$\gamma$, and social temperature $\beta$. The parameters may be
somehow related to the society's political health, such as the
government's characteristics, the relationship between the media and
non-government organizations, or the Internet and the government,
people's understanding of or activity toward democracy, and
society's collective behavior.\\

\noindent {\bf 4. Conclusions}
\hfill\\

It is demonstrated that collective opinion formation emerges in
nonlinear interaction in terms of unstable, periodic or even chaotic
patterns of behavior. The aim of this paper is to formulate a
numerical model of opinion formation or a collective decision-making
procedure using binary and independent strategies. This binary
strategy can be thought of as the two elements yin and yang in
Chinese philosophy, and produces various phenomena in nature on many
specific conditions. Nonlinear interaction can be numerically
realized in terms of social structure, information asymmetries or an
information cascade in a real life social network, complex electoral
processes, decision making processes, etc. A combination of
nonlinear interaction and binary strategy may be the mother of
various political actions or societal behavior.\\

\newpage
\vskip 10mm
\begin{center}
{\bf FIGURE  CAPTIONS}
\end{center}

\vspace {5mm}

\noindent Fig. $1$. Bifurcation map for social temperature $\beta$;
$\alpha=0.2$, $\gamma=0.8$, and $b=3$.
\vspace {10mm}

\noindent Fig. $2$. Plot of $x_{i+1}$ versus $x_i$ in the case of
$\alpha=0.5$, $\gamma=0.5$, and  $b=3$.

\vspace {10mm}

\end{document}